\documentclass[journal]{IEEEtran}

\usepackage{fancyhdr, epsfig, epsf, amsthm, amsmath, amssymb, amsfonts, subfigure, color}
\usepackage{enumerate}
\usepackage{dsfont}
\usepackage[noadjust]{cite}

\usepackage{stfloats}

\newtheorem{proposition}{Proposition}

\def\({\left(}
\def\){\right)}

\def\[{\left[}
\def\]{\right]}
\def\E{\mathsf{E}}

\def\s{\small}
\def\f{\footnotesize}

\newcommand{\chapternote}[1]{%
 \let\thempfn\relax
  \footnotetext[0]{\emph{#1}}
  }

\ifCLASSINFOpdf
\else
\fi
\begin{document}
%
\title{\LARGE Blockchained On-Device Federated Learning }
%
%
%
%

\author{\IEEEauthorblockN{\s Hyesung Kim, Jihong Park$^{\dagger}$, Mehdi Bennis$^{\dagger}$, and Seong-Lyun Kim \vskip -3pt}

\vskip -35pt

\thanks{\vskip -15pt \scriptsize This work was partly supported by Institute of Information \& communications Technology Planning \& Evaluation (IITP) grant funded by the Korea government (MSIT) (No.2018-0-00170, Virtual Presence in Moving Objects through 5G), Basic Science Research Foundation of Korea(NRF) grant funded by the Ministry of Science and ICT (NRF-2017R1A2A2A05069810), and the Mobile Edge Intelligence at Scale (ELLIS) project at the University of Oulu.}
\thanks{\scriptsize H. Kim and S.-L. Kim are with School of Electrical and Electronic Engineering, Yonsei University, Seoul, Korea (email: \{hskim, slkim\}@ramo.yonsei.ac.kr).}
\thanks{\scriptsize$^{\dagger}$J. Park and $^{\dagger}$M. Bennis are with the Centre for Wireless Communications, University of Oulu, 4500 Oulu, Finland (email: \{jihong.park, mehdi.bennis\}@oulu.fi).}
}

\markboth{}%
{Shell \MakeLowercase{\textit{et al.}}: Bare Demo of IEEEtran.cls for IEEE Journals}
%



\maketitle

\begin{abstract}
 By leveraging blockchain, this letter proposes a blockchained federated learning (BlockFL) architecture where local learning model updates are exchanged and verified.  
This enables on-device machine learning without any centralized training data or coordination by utilizing a consensus mechanism in blockchain. 
Moreover, we analyze an end-to-end latency model of BlockFL and characterize the optimal block generation rate by considering communication, computation, and consensus delays.
\end{abstract}

\begin{IEEEkeywords}
On-device machine learning, federated learning, blockchain, latency. 
\end{IEEEkeywords}

%
\IEEEpeerreviewmaketitle

\section{Introduction}

\IEEEPARstart{F}{uture} wireless systems are envisaged to ensure low latency and high reliability anywhere and anytime \cite{PetarURLLC:17,MehdiURLLC:18,UR2Cspaswin:17}. To this end, on-device machine learning is a compelling solution wherein each device stores a high-quality machine learning model and is thereby capable of make decisions, even when it loses connectivity. Training such an on-device machine learning model requires more data samples than each device's local samples, and necessitates sample exchanges with other devices \cite{Jakub_FL16,Brendan17, park_ieee_proceeding}. In this letter, we tackle the problem of training each device's local model by federating with other devices.

One key challenge is that local data samples are owned by each device. {Thus, the exchanges should keep the raw data samples private from other devices. For this purpose, as proposed in Google's federated learning (FL) \cite{Jakub_FL16,Brendan17}, referred to as  {\it vanilla FL}, each device  exchanges its \emph{local model update}, i.e., learning model's weight and gradient parameters, from which the raw data cannot be derived.} As illustrated in Fig.~1-a, the vanilla FL's exchange is enabled by the aid of a central server that aggregates and takes an ensemble average of all the local model updates, yielding a \emph{global model~update}. 
Then, each device downloads the global model update, and computes its next local update until the global model training is completed \cite{Brendan17}. 
Due to these exchanges, the vanilla FL's training completion latency might be tens of minutes or more, as demonstrated in Google's keyboard application \cite{Google_ai2}.

The limitation of the vanilla FL operation is two-fold. Firstly, it relies on a single central server, which is vulnerable to the server's malfunction. This incurs inaccurate global model updates distorting all local model updates. Secondly, it does not reward the local devices. A device having a larger number of data samples contributes more to the global training.
Without providing compensation, such a device is less willing to federate with the other devices possessing few data samples. 

In order to resolve these pressing issues, by leveraging \emph{blockchain} \cite{Bitcoin,info_prop} in lieu of the central server, we propose a \emph{blockchained FL (BlockFL)} architecture, where the blockchain network enables exchanging  devices' local model updates while verifying and providing their corresponding rewards.
 BlockFL overcomes the single point of failure problem and  extends the range of its federation to untrustworthy devices in a public network thanks to a validation process of the local training results.
Moreover, by providing  rewards proportional to the training sample sizes, BlockFL promotes the federation of more devices with a larger number of training samples.

 As shown in Fig.~1-b, the logical structure of BlockFL consists of devices and miners. The miners can physically be either randomly selected devices or separate nodes such as network edges (i.e., base stations in cellular networks), which are relatively free from energy constraints in mining process.
The operation of BlockFL is summarized as follows:
Each device computes and uploads the local model update to its associated miner in the blockchain network;
Miners exchange and verify all the local model updates, and then run the Proof-of-Work (PoW) \cite{Bitcoin};
Once a miner completes the PoW, it generates a block where the verified local model updates are recorded;
and finally, the generated block storing the aggregate local model updates is added to a blockchain, also known as distributed ledger, and is downloaded by  devices. Each device computes the global model update from the new block.

{Note that the global model update of BlockFL is computed locally at each device. A miner's or a device's malfunction does not affect other devices' global model updates.}
For the sake of these benefits, in contrast to the vanilla~FL, BlockFL needs to account for the extra delay incurred by the blockchain network. To address this, the end-to-end latency model of BlockFL is formulated by considering communication, computation, and the PoW delays. The resulting latency is minimized by adjusting the block generation rate, i.e., the PoW difficulty. 



\begin{figure}
\centering
 \includegraphics[width= 0.48\textwidth]{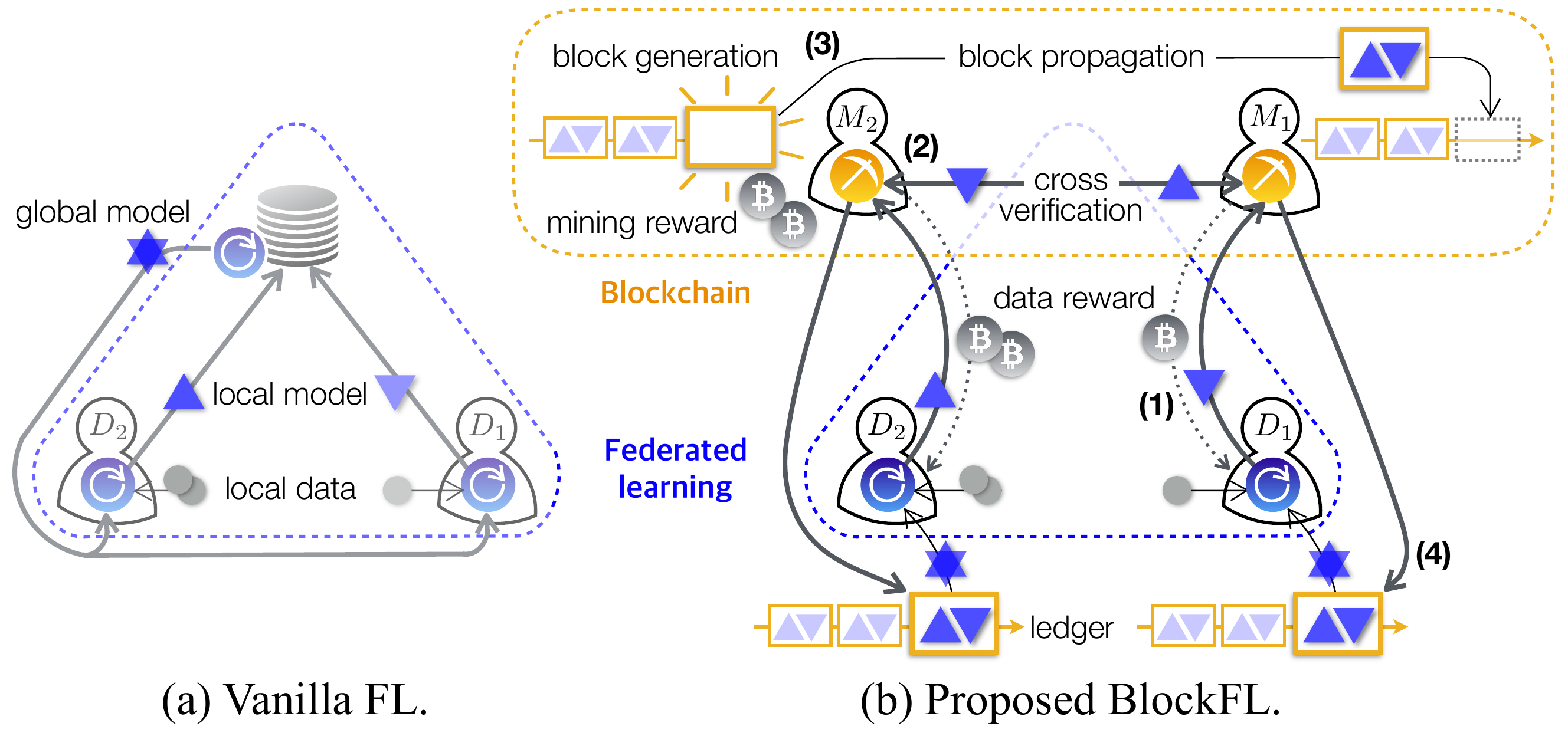}\label{Fig_FLBC}\vskip -10pt
\caption{An illustration of (a) the vanilla federated learning (FL)~\cite{Jakub_FL16,Brendan17} and (b) the proposed blockchained FL (BlockFL) architectures.}
\end{figure}

\section{Architecture and Operation}





\textbf{FL operation in BlockFL:} The FL under study is operated by a set of devices {\f$\mathcal{D}=\{1,2,\cdots,N_D\}$} with {\f$|\mathcal{D}|=N_D$}. The $i$-th device~$D_i$ owns a set of data samples $\mathcal{S}_i$ with {\f$|\mathcal{S}_i|=N_i$}, and trains its local model. The local model updates of the device $D_i$ is uploaded to its associated miner $M_j$ that is uniformly randomly selected out of a set of miners {\f$\mathcal{M}\!=\!\{1,2,\cdots,\!N\!_M\}$}.

Our distributed model training focuses on solving a regression problem in a parallel manner, considering a set of the entire devices' data samples $\mathcal{S}=\cup_{i=1}^{N_D} \mathcal{S}_i$ with $|\mathcal{S}|=N_S$. The $k$-th data sample $s_k\in\mathcal{S}$ is given as $s_k=\{x_k,y_k\}$ for a $d$-dimensional column vector $x_k\in\mathbb{R}^d$ and a scalar value $y_k\in\mathbb{R}$. The objective is to minimize a loss function {\s$f(w)$} for a global weight vector $w\in\mathbb{R}^d$. The loss function {\s$f(w)$} is chosen as the mean squared error:
$
f(w) \!=\! \frac{1}{N_S} \sum_{i=1}^{N_D}\sum_{s_k\in\mathcal{S}_i} f_k(w), $
\noindent where {\s$f_k(w)\!=\!(x_k^\top w - y_k)^2\!/2$}.  Other loss functions under deep neural networks can readily be incorporated, as done in~\cite{FL_v2x}.


In order to solve the above problem, following the vanilla FL settings in \cite{Jakub_FL16}, {the model of the device $D_i$ is locally trained
via a stochastic variance reduced gradient algorithm \cite{Jakub_FL16}, and all devices' local model updates are aggregated using a distributed approximate Newton method.}
For each epoch, the device $D_i$'s local model is updated with the number $N_i$ of iterations. 
At the $t$-th local iteration of the $\ell$-th epoch, the local weight {\s$w_i^{(t,\ell)}\in\mathbb{R}^d$} is:
\vspace{-10pt}

{{\f\begin{align} \label{Eq:wlocal}
w_i^{(t,\ell)}\! = w_i^{(t-1,\ell)}\!-\! \frac{\beta}{N_i}\!\(\! \[\nabla f_k(w_i^{(t-1,\ell)}) \!-\! \nabla f_k (w^{(\ell)} )\] \!\!+\!\! \nabla f(w^{(\ell)}\!)\!\!\)\!,
\end{align}}\normalsize }
\noindent where {\s$\beta>0$} is a step size, {\s$w^{(\ell)}$} indicates the global weight at the $\ell$-th epoch, and {\s$\nabla f(w^{(\ell)})={1}/{N_S}\cdot \sum_{i=1}^{N_D}\sum_{s_k\in\mathcal{S}_i} \nabla f_k(w^{(\ell)})$}. Let {\s$w_i^{(\ell)}$} denote the local weight after the last local iteration of the $\ell$-th epoch, i.e., {\s$w_i^{(\ell)} \!=\! w_i^{(N_i,\ell)}\!$}. Then, 
\begin{align}\label{Eq:wglobal}
{\f{w^{(\ell)}=w^{(\ell-1)}+\sum_{i=1}^{N_D} \frac{N_i}{N_S} \( w_i^{(\ell)}-w^{(\ell-1)} \).}}
\end{align} \normalsize

In  vanilla FL in \cite{Brendan17,Jakub_FL16}, the device $D_i$ uploads its local model update {\f$(w_i^{(\ell)}, \{\nabla f_k(w^{(\ell)}) \}_{s_k\in\mathcal{S}_i})$} to the central server, with the model update size {\s$\delta_\text{m}$} that is identically given for all devices. The global model update {\f$(w^{(\ell)}, \nabla f(w^{(\ell)}) )$} is computed by the server. {\it In BlockFL, the server entity is substituted with a blockchain network} as detailed in the following description.

\textbf{Blockchain operation in BlockFL:} In the  BlockFL, the blocks and their verification by the miners in $\mathcal{M}$ are designed so as to exchange the local model updates truthfully through a distributed ledger. Each block in a ledger is divided into its body and header parts \cite{Bitcoin}. 
In BlockFL, the body stores the local model updates of the devices in $\mathcal{D}$, i.e., {\f$(w_i^{(\ell)}, \{\nabla f_k(w^{(\ell)}) \}_{s_k\in\mathcal{S}_i})$} for the device $D_i$ at the $\ell$-th epoch, as well as its local computation time {\s$T_{\text{local},i}^{(\ell)}$} that is discussed at the end of this subsection. 
The header contains the information of a pointer to the previous block, block generation rate $\lambda$, and the output value of the PoW. The size of each block is set as {\s$h+ \delta_\text{m} N_D$}, where $h$ and {\s$\delta_\text{m}$} are the header and model update sizes, respectively.
Each miner has a candidate block that is filled with the local model updates from its associated devices and/or other miners. The {filling procedure}  continues until it reaches the block size or a  waiting time $T_\text{wait}$.

\begin{figure}
\centering
\includegraphics[width=0.47\textwidth]{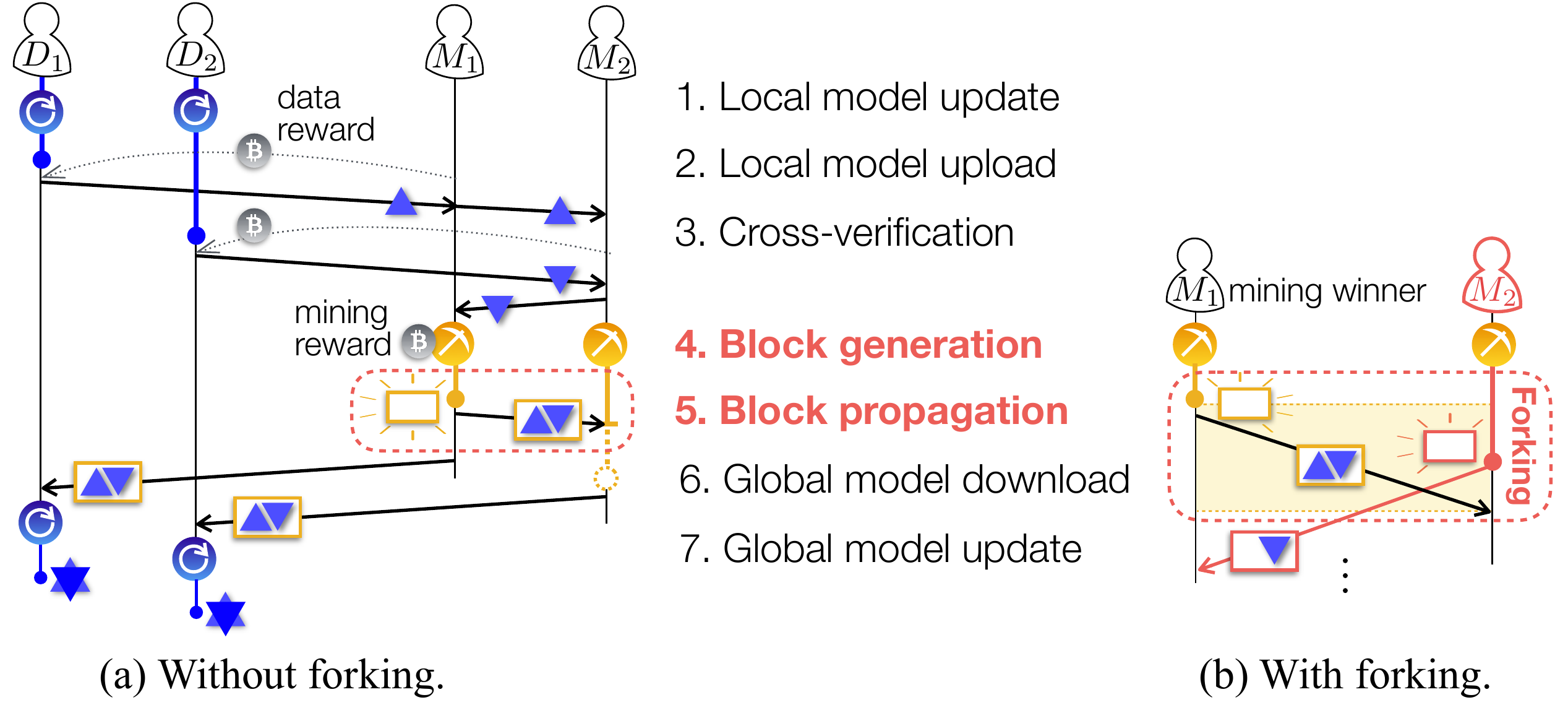}   
\vskip -11pt\caption{{\f{The one-epoch operation of BlockFL with and without forking.}}}\label{protocol} 
\end{figure}

{ Afterwards, following the PoW \cite{Bitcoin}, the miner randomly generates a hash value by changing its input number, i.e., nonce, until the generated hash value becomes smaller than a target value. Once the miner $M_1$ succeeds in finding the hash value, its candidate block is allowed to be a new block as shown in Fig 2. Here, the block generation rate $\lambda$  can be controlled by the PoW difficulty, e.g., the lower PoW target hash value, the smaller~$\lambda$.} 
{Due to simpleness and robustness, the PoW is  applied to the wireless systems as in \cite{Petar_iot_BC18, Niyato18_ICC}. 
BlockFL can also use other consensus algorithms such as the proof-of-stake (PoS) or Byzantine-fault-tolerance (BFT), which may require more complicated operation and preliminaries to reach consensus among miners.} 

The generated block is propagated to all other miners. 
To this end, as done in \cite{Bitcoin}, all the miners receiving the generated block are forced to stop their PoW operations and to add the generated block to their local ledgers. As illustrated in Fig~2, if another miner $M_2$ succeeds in its block generation within the propagation delay of the firstly generated block, then some miners may mistakenly add this secondly generated block to their local ledgers, known as \emph{forking}. In BlockFL, forking makes some devices apply an incorrect global model update to their next local model updates. Forking frequency increases with  $\lambda$ and the block propagation delay, and its mitigation incurs an extra delay, to be elaborated in Sect.~III.

The blockchain network also provides rewards for data samples to the devices and for the verification process to the miners, referred to as \emph{data~reward} and \emph{mining reward}, respectively. The data reward of the device $D_i$ is received from its associated miner, and its amount is proportional to the data sample size $N_i$. When the miner $M_j$ generates a block, its mining reward is earned by the blockchain network, as done in the conventional blockchain structure \cite{Bitcoin}. The amount of mining reward is proportional to the aggregate data sample size of its all associating devices, namely, {\f$\sum_{i=1}^{N_{M_j}} N_i$} where $N_{M_j}$ denotes the number of  devices associated with the miner $M_j$. 
{It is noted that the BlockFL can further be improved using a reward mechanism, which considers not only the size but also the quality of data sample that affects the accuracy of FL. }      
Untruthful devices may inflate their sample sizes with arbitrary local model updates.
Miners verify truthful local updates before storing them by comparing the sample size $N_i$ with its corresponding computation time {\f$T_{\text{local},i}^{(\ell)}$}. This can be guaranteed in practice by Intel's software guard extensions,
  allowing applications to be operated within a protected environment, which is utilized in  blockchain technologies~\cite{PoET}.


\textbf{One-epoch BlockFL operation:} As depicted in Fig.~\ref{protocol}, the BlockFL operation of the device $D_i$ at the $\ell$-th epoch is described by the following seven steps. 


{\s\noindent	\textsf{1.} \textsf{Local model update}}: 
	The device $D_i$ computes \eqref{Eq:wlocal} with the number $N_i$ of iterations.
	
{\s\noindent	\textsf{2.} \textsf{Local model upload}}: 
	The device $D_i$ uniformly randomly associates with the miner $M_i$; if $\mathcal{M}=\mathcal{D}$, then $M_i$ is selected from $\mathcal{M}\backslash D_i$. The device uploads the local model updates {\f$(w_i^{(\ell)}, \{\nabla f_k(w^{(\ell)}) \}_{s_k\in\mathcal{S}_i})$} and the corresponding local computation time {\s$T_{\text{local},i}^{(\ell)}$} to the associated miner.

	{\s\noindent \textsf{3.} \textsf{Cross-verification}}: 
	Miners broadcast the obtained local model updates. At the same time, the miners verify the received local model updates from their associated devices or the other miners.
%
%
	The verified local model updates are recorded in the miner's candidate block, until its reaching the block size {\s$(h+\delta_\text{m}N_D)$} or the maximum waiting time {\s$T_\text{wait}$}.

{\s\noindent	\textsf{4.} \textsf{Block generation}}: 
	Each miner starts running the PoW until either it finds the nonce or it receives a generated block.
	
	{\s \noindent\textsf{5.} \textsf{Block propagation}}: 
{	Denoting as $M_{\hat{o}}\in\mathcal{M}$ the miner who first finds the nonce. 
	Its candidate block is generated as a new block and broadcasted to all miners. In order to avoid forking, an ACK, including whether forking occurs or not, is transmitted once each miner receives the new block. If a forking event occurs, the operation restarts from {\s\textsf{Step 1}}. A miner that generates a new block waits until a predefined maximum block ACK waiting time ${\footnotesize T_{{\text{a,wait}}}}$. }

\noindent	{\s\textsf{6.} \textsf{Global model download}}: 
	The device $D_i$ downloads the generated block from its associated miner.

\noindent	{\s\textsf{7.} \textsf{Global model update}}: 
	The device $D_i$ locally computes the global model update in \eqref{Eq:wglobal} by using the aggregate local model updates in the generated block.

\noindent The procedure continues until 
satisfying {\f$|w^{(L)}\!-\!w^{(L\!-\!1)}|\!\leq \!\varepsilon$}. 
%

{Centralized FL  is vulnerable to the server's malfunction that distorts all devices' global models.
However, in  BlockFL, the global model update is computed locally at each device, which is  robust against the malfunction and prevents excessive computational overheads of~miners. }

 \section{ End-to-End Latency Analysis}

We investigate the optimal block generation rate {\s$\lambda^*$} minimizing learning completion latency {\f$T_o$}, defined as the total time during $L$ epochs at a  randomly selected device {\f$D_o\!\in\!\mathcal{D}$}.

 \textbf{One-epoch BlockFL latency model:} 
The $\ell$-th epoch latency {\s$T_o^{(\ell)}$} is determined by computation, communication, and block generation delays.
First, computation delays are brought by {\s\textsf{Steps 1}} and~{\s\textsf{7}} in Sect. II. Let $\delta_\text{d}$ denote a single data sample's size  identical for all data samples. {Processing $\delta_\text{d}$ with the clock speed $f_c$ requires $\delta_\text{d}/f_c$. Local model updating delay {\s$T_{\text{local},o}^{(\ell)}$} in {\s\textsf{Step 1}} is thus given as {\s$T^{(\ell)}_{\text{local},o}=\delta_\text{d} N_o/f_c$}. Likewise, global model updating delay {\s$T_{\text{global},o}^{(\ell)}$} in {\s\textsf{Step 7}} is evaluated as {\s$T_{\text{global},o}^{(\ell)} = \delta_\text{m}N_D/f_c$}. Note that $\delta_d$ and $\delta_m$ change with applications types.}
Second, communication delays are entailed by {\s\textsf{Steps~2}} and~{\s\textsf{6}} between devices and miners. Measuring the achievable rate under additive white Gaussian noise channels, local model uploading delay {\s$T_{\text{up},o}^{(\ell)}$} in {\s\textsf{Step~2}} is computed as {\s$T_{\text{up},o}^{(\ell)} \!=\! \delta_\text{m}/[W_\text{up}\log_2(1 \!+\! \gamma_{\text{up},o})]$}, where {\s$W_\text{up}$} is the uplink bandwidth allocation per device and {\s$\gamma_{\text{up},o}$} is the miner $M_o$'s received signal-to-noise ratio (SNR). The global model downloading delay {\s$T_{\text{dn},o}^{(\ell)}$} in {\s\textsf{Step 6}} is given as {\f$T_{\text{dn},o}^{(\ell)} \!=\! (h \!+\! \delta_\text{m} N_D)/[W_\text{dn}\log_2(1\! +\! \gamma_{\text{dn},o})]$}, where {\s$W_\text{dn}$} is the downlink bandwidth  per device and {\s$\gamma_{\text{dn},o}$} is the device $D_o$'s  SNR.

For {\s\textsf{Steps~3}} and~{\s\textsf{5}}, assuming verification processing time is negligible compared to the communication delays, cross-verification delay~{\s$T_{\text{cross},o}^{(\ell)}$} in {\s\textsf{Step~3}} is {\f$T_{\text{cross},o}^{(\ell)}\!\!=\!\! \max\{ T_\text{wait}- (T_{\text{local},o}^{(\ell)}\!+\! T_{\text{up},o}^{(\ell)}), \sum_{M_{j}\in\mathcal{M}\backslash M_o} \delta_\text{m} N_{M_j} /[W_\text{m}\log_2(1+\gamma_{oj})] \} $} under frequency division multiple access (FDMA), where {\s$W_\text{m}$} is the bandwidth allocation per each miner link and {\s$\gamma_{oj}$} is the miner $M_j$'s received SNR from the miner $M_o$. Denoting as $M_{\hat{o}}\in\mathcal{M}$ the miner who first finds nonce, referred to as the mining winner, {total block propagation delay $T_{\text{bp},\hat{o} }^{(\ell)}$ in {\s\textsf{Step~5}} is given as~{\s$T_{\text{bp},\hat{o} }^{(\ell)}\!=\!\max_{M_j\in\mathcal{M}\backslash M_{\hat{o}}} \{t_{\text{bp},j}^{(\ell)},\footnotesize T_{{\text{a,wait}}}\}$} under FDMA.}~The term {\s$t_{\text{bp},j}^{(\ell)}\!=\!(h+\delta_\text{m} N_{D}) /[W_\text{m}\log_2(1 + \gamma_{\hat{o}j})]$} represents the block propagation delay from the mining winner $M_{\hat{o}}$ to $M_{j}\in\mathcal{M}\backslash M_{\hat{o}}$, and {\s$\gamma_{\hat{o}j}$} is the miner $M_j$'s received SNR.
{Lastly, in {\s\textsf{Step 4}}, block generation delay {\s$T_{\text{bg},j}^{(\ell)}$} of the miner $M_j\in\mathcal{M}$ follows an exponential distribution with mean~{\s$1/\lambda$}, as modeled in \cite{info_prop}.} The delay of interest is the mining winner $M_{\hat{o}}$'s block generation delay {\s$T_{\text{bg},\hat{o}}^{(\ell)}$}. Finally, the $\ell$-th epoch latency {\s$T^{(\ell)}_o$} is

\vspace{-2pt}\f\begin{align} 
\!\!\!T_o^{(\ell)} \!\!= \! \!N_\text{fork}^{(\ell)}\( T_{\text{local},o}^{(\ell)} \!+\! T_{\text{up},o}^{(\ell)} \!+\! T_{\text{cross},o}^{(\ell)}\! \!+\! T_{\text{bg},\hat{o}}^{(\ell)} \!+\! T_{\text{bp},\hat{o}}^{(\ell)}\)\! \!+\! T_{\text{dn},o}^{(\ell)} \!+\! T_{\text{global},o}^{(\ell)}, \label{Eq:E2Elatency}
\end{align}\normalsize 
\noindent where {\s$N_\text{fork}^{(\ell)}$} denotes the number of forking occurrences in the $\ell$-th epoch, and follows a geometric distribution with mean {\s$1/(1-p_\text{fork}^{(\ell)})$}, with the forking probability $p_\text{fork}^{(\ell)}$ at the $\ell$-th epoch. Following {\s\textsf{Step~5}}, the forking probability is represented as:

{\f\begin{align}
p_\text{fork}^{(\ell)} =1-\prod_{M_j\in\mathcal{M}\backslash M_{\hat{o} }} \Pr\(t_j^{(\ell)} - t_{\hat{o}}^{(\ell)} >  t_{\text{bp},j}^{(\ell)}  \),  \label{p_f} 
\end{align}} 
\vskip -5pt

\noindent where the term {\f$t_j^{(\ell)} = T_{\text{local},j}^{(\ell)} + T_{\text{up},j}^{(\ell)} + T_{\text{cross},j}^{(\ell)}+ T_{\text{bg},j}^{(\ell)} $} is the cumulated delay until the miner $M_j$ generates a block.

\textbf{Latency optimal block generation rate:}
Using the one-epoch latency \eqref{Eq:E2Elatency}, we derive the optimal block generation rate $\lambda^*$ that minimizes the $\ell$-th epoch latency averaged over the PoW process. Here, the PoW process affects the block generation delay {\s$T_{\text{bg},\hat{o}}^{(\ell)}$}, block propagation delay {\s$T_{\text{bp},\hat{o}}^{(\ell)}$}, and the number {\s$N_{\text{fork}}^{(\ell)}$} of forking occurrences, which are inter-dependent due to the mining winner $M_{\hat{o}}$. 
We consider the case where all miners synchronously start their PoW processes by adjusting {\s$T_{\text{wait}}$} such that {\s$T_{\text{cross},o}^{(\ell)}\!=\!T_\text{wait}\!- \!(T_{\text{local},o}^{(\ell)}\!+\!T_{\text{up},o}^{(\ell)})$}. In this case, even the miners completing the cross-verification earlier wait until $T_{\text{wait}}$,  providing the latency upper bound. With this  approximation, we derive the optimal block generation rate ${\lambda}^*$ as follows.
\vspace{-5pt}\begin{proposition} \emph{With the PoW synchronous approximation, i.e., {\s$T_{\text{cross},o}^{(\ell)}=T_\text{wait}- (T_{\text{local},o}^{(\ell)}+T_{\text{up},o}^{(\ell)})$}, the block generation rate ${\lambda}^*$ minimizing the $\ell$-th epoch latency {\f$\mathbb{E}[T^{(\ell)}_o]$} averaged over the PoW process is given by:}
\vskip -12pt
{\f\emph{\begin{align}
{{{\lambda}^*\approx {2}{\(  T^{(\ell)}_{\text{bp},\hat{o} } \[1+\sqrt{1+ {4N_M\(1 + T_{\text{wait}} / T^{(\ell)}_{\text{bp},\hat{o} }\)}} \] \)^{-1}}.}} \nonumber
\end{align}}\normalsize} 
\vskip -10pt
{
{\noindent Proof: 
\emph{
Applying the synchronous PoW approximation and the mean {\f$1/(1-p_\text{fork}^{(\ell)})$} of the geometrically distributed {\f$N_{\text{fork}}^{(\ell)}$} to \eqref{Eq:E2Elatency},} 

\vskip -5pt
{\f{\begin{align} 
\E[T^{(\ell)}_o] \approx \( T_{\text{wait}}+  \E[T_{\text{bg},\hat{o}}^{(\ell)}] \) /  \(1- p_\text{fork}^{(\ell)}\) + T_{\text{dn},o}^{(\ell)} + T_{\text{global},o}^{(\ell)}. \label{E_T_o1}
\end{align}}}}
\noindent \emph{The terms {\f$T_\text{wait}$}, {\f$T_{\text{dn},o}^{(\ell)}$}, {\f$T_{\text{global},o}^{(\ell)}$} are constant delays given in Sect.~II.} 
\emph{For the probability {\f$p_\text{fork}^{(\ell)}$}, using \eqref{p_f} with {\f$t_j^{(\ell)}
\!-\!t_{\hat{o}}^{(\ell)}\! =\! T_{\text{bg},j}^{(\ell)}\!-\! T_{\text{bg},\hat{o}}^{(\ell)}$} under the synchronous approximation, we obtain {\f$p_\text{fork}^{(\ell)}$} as:}
{\f\emph{$
p_\text{fork}^{(\ell)} = 1-e^{\lambda \sum_{M_j\in\mathcal{M}\backslash M_{\hat{o} }}T^{(\ell)}_{\text{bp},j}}, \label{p_f2}
$}}
\noindent\emph{where {\f$T^{(\ell)}_{\text{bp},j}$} is a constant delay given in Sect. II-A. Next, for the delay {\f$\E[T^{(\ell)}_{\text{bg},\hat{o}}]$}, using the definition of {\f$T^{(\ell)}_{\text{bg},\hat{o}}$} and the complementary cumulative distribution function (CCDF) of the exponentially distributed {\f$T_{\text{bg},j}^{(\ell)}$}, we derive {\f$T^{(\ell)}_{\text{bg},\hat{o}}$}'s CCDF as: ${\f
\text{Pr}\Big(T^{(\ell)}_{\text{bg},\hat{o}} >x \Big) \!=\! \prod_{j=1}^{N\!_M}\text{Pr}\Big(  T^{(\ell)}_{\text{bg},j} > x \Big)\!=\! e^{-\lambda N_M x }.
}$ Applying the total probability theorem yields {\f$\E[T^{(\ell)}_{\text{bg},\hat{o}}]\!=\!1/(\lambda N_M)$}. Finally, combining all these terms, \eqref{E_T_o1} is recast as: {\f$\E[T^{(\ell)}_o]\!\approx\! \(T_{\text {wait}} + {1}/{\lambda N_M}\)\!e^{\lambda \sum_{M_j\in\mathcal{M}\backslash M_{\hat{o} }}T^{(\ell)}_{\text{bp},j}}\! +\! T_{\text{dn},o}^{(\ell)} \!+\! T_{\text{global},o}^{(\ell)},$} which is convex for $\lambda$. The optimum $\lambda^*$ is thus directly derived.  \hfill$\blacksquare$}
}
\end{proposition}\vskip -7pt


{For a larger $\lambda$, 
the forking event occurs more frequently, increasing the learning completion latency. 
On the contrary, for a high PoW difficulty with a low $\lambda$, the block generation time incurs its own overheads  with excessive latency. 
}

\section{Numerical Results and Discussion}

We numerically evaluate the proposed blockFL's average learning completion latency {\s$\E[T_o] \!=\! \sum_{\ell=1}^L \E[T_o^{(\ell)}]$}. By default, we consider {\s$N_D=N_M=10$}, and {\s$N_i\!\sim\!\text{Uni}(10,50)\;\forall D_i\!\in\!\mathcal{D}$}. Following the 3GPP LTE Cat. M1 specification, we use {\s$W_\text{up}\!=\!W_\text{dn}\!=\!W_\text{m}\!=\!300$} KHz and {\s$\gamma_{\text{up},o}\!=\!\gamma_{\text{dn},o}\!=\!\gamma_{oj}\!=\!10$} dB. Other simulation parameters are given as: {\s$\delta_\text{d}\!=\!100$} Kbit, {\s$\delta_\text{m}\!=\!5$} Kbit, {\s$h\!=\!200$} Kbit, {\s$f_c\! =\! 1$}~GHz, {\s$T_\text{wait}\!=\! 50$} ms, {\footnotesize $T_{{\text{a,wait}}}\!=\!500$} ms.


\begin{figure}
\centering
\subfigure[\vspace{-10pt}\f{With respect to {\f$\lambda$}}.]{\includegraphics[width=.42\columnwidth]{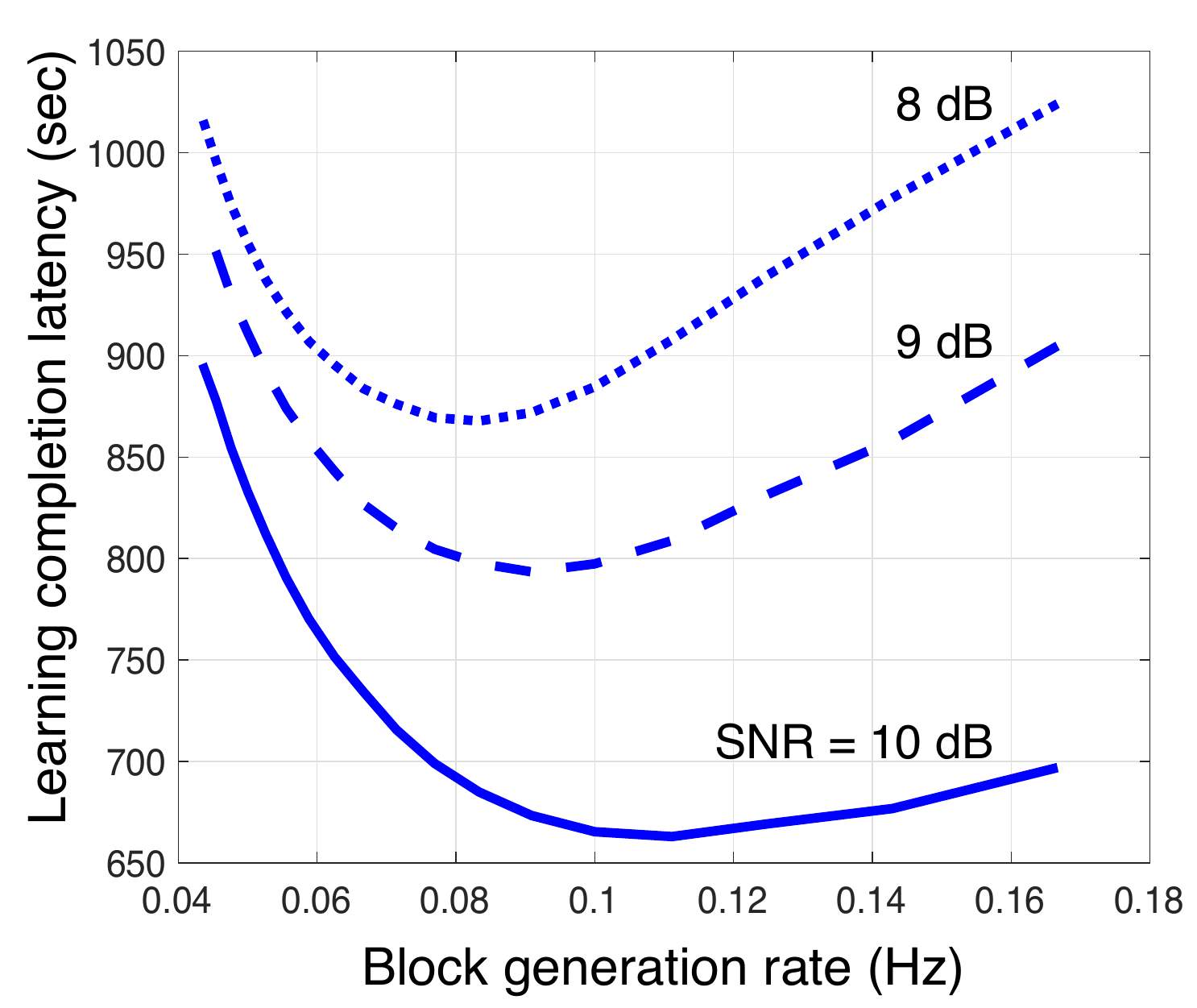}}  
\subfigure[\vspace{-10pt}\f{Test accuracy.}]{\includegraphics[width=0.44\columnwidth]{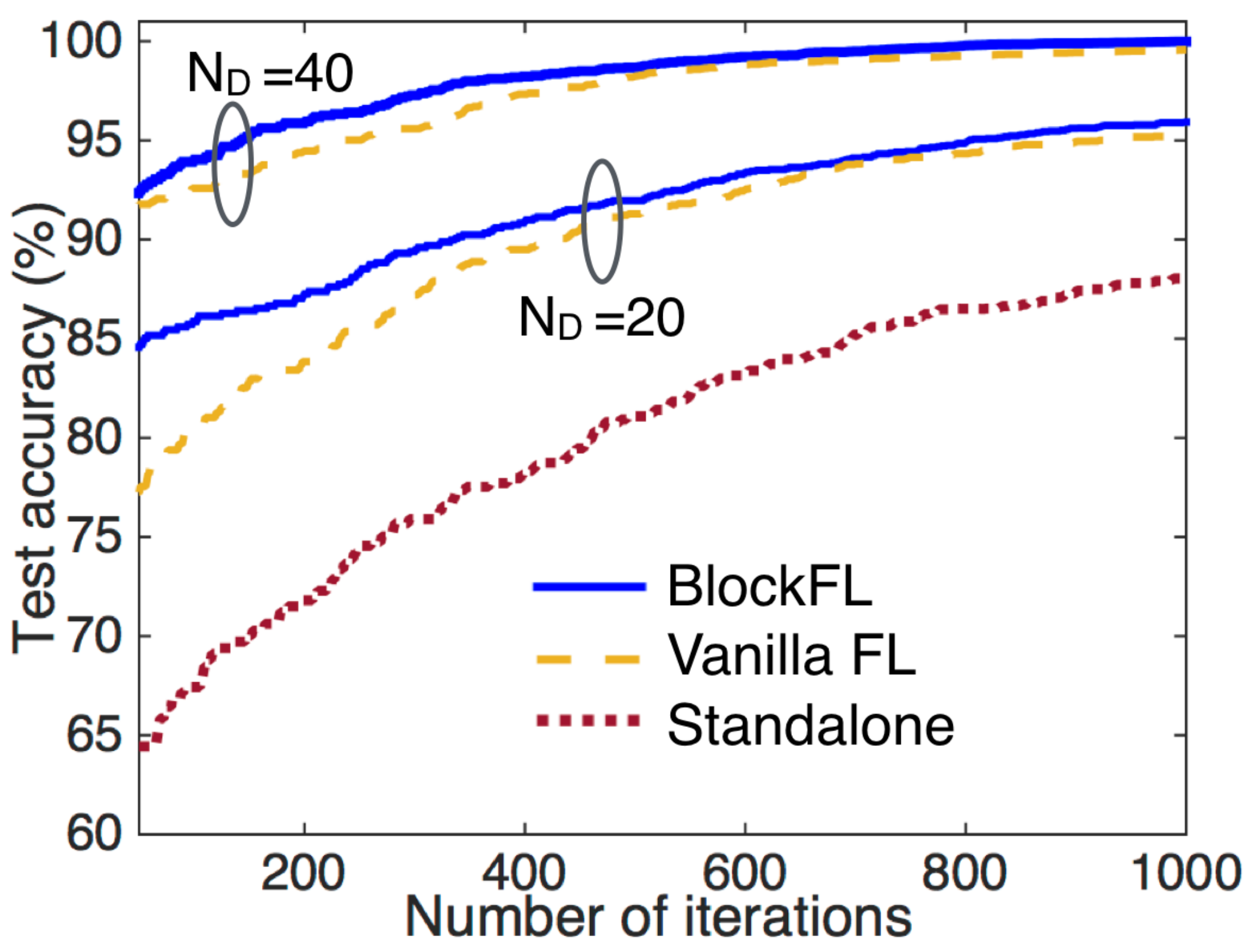}} 
 \vskip -10pt
\caption{{Average learning completion latency (a) versus block generation rate~$\lambda$ and (b) test accuracy of BlockFL, Vanilla FL, and standalone without federation  ({\f$\gamma_{\text{up},o}=\gamma_{\text{dn},o}=\gamma_{oj}=\textsf{SNR}$}).}} \label{fig3} 
\end{figure}


Fig.~3-a shows the impact of block generation rate {\s$\lambda$} on the BlockFL's average learning completion latency. In Fig.~3-a, we observe that the latency is convex-shaped and is decreasing with the SNRs. For the optimal block generation rate {\s$\lambda^*$}, the minimized average learning completion latency obtained from Proposition~1 is always longer by up to $1.5$\% than the simulated minimum latency.
{In Fig.~{3-b}, the BlockFL and the vanilla FL  achieve  almost the same accuracy for an identical $N_D$. On the other hand, the learning completion latency of our BlockFL is lower than that of the vanilla FL ($N_M\!=\!1$) as in Fig.~4-a, which
 shows the scalability in terms of the numbers {\s$N_M$} and {\s$N_D$} of miners and devices, respectively.} The average learning completion latency is computed for {\s$N_M\!=\!1,10$} with or without the miners' malfunction. The malfunction is captured by adding Gaussian noise {\s$\mathcal{N}(-0.1,0.01)$} to each miner's aggregate local model updates with probability $0.05$. Without malfunction, a larger {\s$N_M$} increases the latency due to the increase in their cross-verification and block propagation delays. 
In BlockFL, each miner's malfunction only distorts its associated device's global model update. Such distortion can be restored by federating with other devices that associate with the miners operating normally. Hence, a larger {\s$N_M$} may achieve a shorter latency for {\s$N_M\!=\!10$} with the malfunction.

Fig.~4-a shows that there exists a latency-optimal number {\s$N_D$} of devices. {A larger {\s$N_D$} enables to utilize a larger amount of data samples, 
whereas it increases each block size and block exchange delays, resulting in the convex-shaped latency.} 

{In Fig.~4-b, we assume that some  miners cannot participate if their battery level is lower than a predefined threshold value $\theta_e$, a normalized battery level, $\theta_e \!\in\! [0,\!1]$. 
Without the malfunction of miner nodes,  the learning completion latency becomes larger for a lower $\theta_e$ due to an increase in cross-verification and block propagation delays. On the contrary, when the malfunction occurs,  a lower $\theta_e$ achieves a shorter latency because more miners federate with leading to robust global model updates. 
Fig. 4-c  shows the overtake probability
that a malicious miner will ever form a new blockchain whose length is longer than a blockchain formed by honest miners. 
The overtake probability goes to zero if just a few blocks have already been chained by honest miners. 
Although the malicious miner begins the first PoW with the honest miners, the larger number of miners prevents the overtake.}

\begin{figure}
\centering
\includegraphics[width=0.485\textwidth]{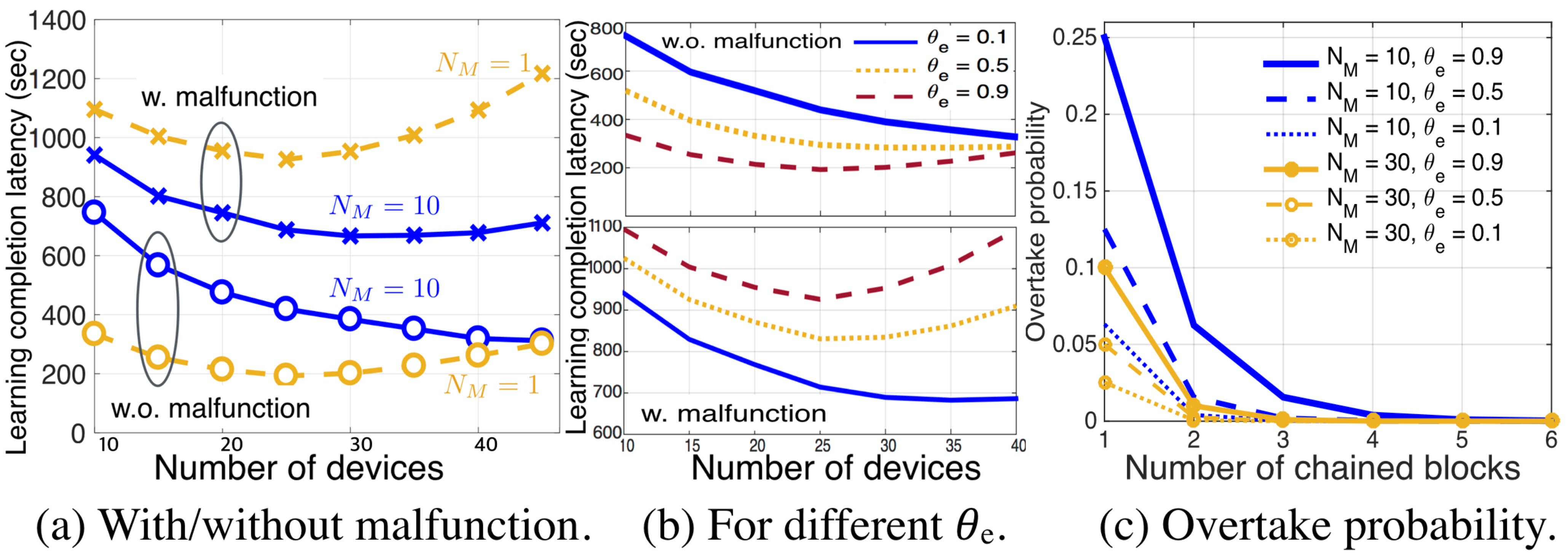}   
\vskip -12pt\caption{{\f{Average learning completion latency versus the number of devices, (a) under the miners' malfunction,  {(b) for different energy constraints $\theta_e$, and (c) overtake probability with respect to the number of chained blocks.}}}}\label{fig45} 
\end{figure}

\bibliographystyle{ieeetr}  

\bibliography{BC_FL}

\end{document}